\documentclass[aps,twocolumn,showpacs,showkeys,floatfix]{revtex4}
\usepackage{amsfonts}
\usepackage{amsmath}
\usepackage{amssymb}
\usepackage{graphicx}

\begin{document}
\title{Electric field induced hyperfine level-crossings in ($n$D)Cs at two-step laser excitation: experiment and theory}

\author{M. Auzinsh, K. Blushs, R. Ferber, F. Gahbauer, A. Jarmola, and M. Tamanis}

\affiliation{Department of Physics and Institute of Atomic Physics and Spectroscopy, University of Latvia, 19 Rainis Blvd., LV-1586 Riga, Latvia}

\begin{abstract}
The pure electric field level-crossing of $m_F$ Zeeman sublevels of hyperfine $F$ levels at
two-step laser excitation was described theoretically and studied experimentally for the $n$D$_{3/2}$
states in Cs with $n$ = 7, 9 and 10, by applying a diode laser in the first 6S$_{1/2} \rightarrow$ 6P$_{3/2}$
 step and a diode or dye laser for the second 6P$_{3/2} \rightarrow n$D$_{3/2}$ step. Level-crossing
resonance signals were observed in the $n$D$_{3/2} \rightarrow$ 6P$_{1/2}$ fluorescence. A theoretical
model was developed to describe quantitatively the resonance signals by correlation analysis of the
optical Bloch equations in the case
when an atom simultaneously interacts with two laser
fields in the presence of an external dc electric field. The simulations described well the
experimental signals. The tensor polarizabilities $\alpha_2$ (in $a_0^3$) were determined to be
$7.45(20)\times10^4$ for the 7D$_{3/2}$ state and $1.183(35) \times 10^6$ for the 9D$_{3/2}$ state; a well established
$\alpha_2$ value for 10D$_{3/2}$ was used to calibrate the electric field. The $\alpha_2$
value for the 7D$_{3/2}$ state differed by ca. 15\% from the existing experimentally measured
value.

\end{abstract}

\date{October 31, 2005}


\maketitle

\section{Introduction}
The first experimental demonstration of crossings of certain magnetic ($m_F$)
components of hyperfine structure (hfs) levels $F$ at non-zero electric field ${\bf \mathcal{E}}$ was reported
in 1966 in a paper by Khadjavi, Happer and Lurio \cite{Khadjavi66}.  In this work,
the authors observed the Stark effect in the second excited state of the
alkali metal atoms $^{85,87}$Rb (6P$_{3/2}$) and $^{133}$Cs (7P$_{3/2}$).
Using resonant excitation from a gas-discharge lamp, they observed resonances
at the level-crossing positions in
the fluorescence signals from single-step broad-line light excitation.  Such
a method of Stark level-crossing spectroscopy was applied by
the same authors to determine experimentally the
tensor polarizabilities $\alpha_{2}$ in these states, as well as to determine
$\alpha_2$ in $^{39}$K(5P$_{3/2}$) \cite{Khadjavi68,Schmieder71}. Later on, however,
it became more popular to vary the magnetic field in the presence of a constant
electric field.  This way of inducing level-crossings was preferred, perhaps
because  magnetic fields were
easier to produce and control. Such techniques were used to
measure the tensor polarizabilities $\alpha_2$ of alkali atoms
by Svanberg and co-authors (see
\cite{Svanberg72,Belin75,Belin76a,Belin76b}
and references
therein). In particular, this method was used in a two-step excitation with a radio-frequency
discharge lamp and a narrow-width dye laser \cite{Belin75,Belin76a,Belin76b}.
The development of
narrow line-width lasers enabled Stark shifts to be measured directly
\cite{Svanberg73} by scanning
the electric field at a fixed laser frequency.  Both methods made it possible to
determine a large number of excited S and D state scalar and tensor
polarizabilities of Rb and Cs, achieving an  accuracy of some 5\%
(see \cite{Fredrikson77} for a review). The use of electro-optically modulated laser radiation
allowed Xia and
co-authors \cite{Xia97} to measure the scalar and tensor polarizabilities of
$(10-13)$D$_{3/2,5/2}$  states of Cs with an accuracy better than 0.3\%,
which is better than for any other atomic state.

Thus, to our knowledge there has been no experimental observation of
purely electric field level-crossing resonances of $m_F$ hfs levels at two-step,
or any multi-step, laser excitation.  At the same time, no detailed theoretical
descriptions of the expected signals at two-step excitation
had been reported in the literature until now.

The purpose of the present investigation was (i) to observe the Stark
effect induced level-crossing resonances at two-step laser excitation; (ii) to
develop a proper theoretical description based on the optical Bloch equations
for radiation fields with finite spectral widths \cite{Blush04};
and (iii) to study experimentally
the tensor polarizabilities $\alpha_2$ for the $n$D$_{3/2}$ states of Cs
atoms which were accessible with laser sources at our disposal, namely
the states with $n=7,9,$ and $10$.  The 7D$_{3/2}$ state was of particular interest because of a
considerable discrepancy between the only known measured value for the
polarizability $\alpha_2$ \cite{Wessel87}
and its theoretical estimate given in \cite{Wijngaarden94}.
At the same time, the $\alpha_2$ value for the 10D$_{3/2}$ Cs state was measured
with unprecedented accuracy of 0.1\% \cite{Xia97} and remarkably agreed, within
0.25\%, with its calculated counterpart in \cite{Wijngaarden94}.  As a result,
we used the signal from the 10D$_{3/2}$ state to calibrate the
electric field produced in our Cs cell.  A measurement of both
states in the same experimental arrangement allowed us to measure the
polarizability in the 7D$_{3/2}$ state as well as in the
9D$_{3/2}$ state with greater confidence.
Furthermore, Stark effect studies in highly excited Cs states were particularly
interesting, as Cs might be useful as a tracer gas to image
electric fields \cite{Auzinsh01} at room temperature, or even lower temperatures.

\section{Experiment}
\subsection{Method}
\begin{figure}
\includegraphics[width=7.5cm]{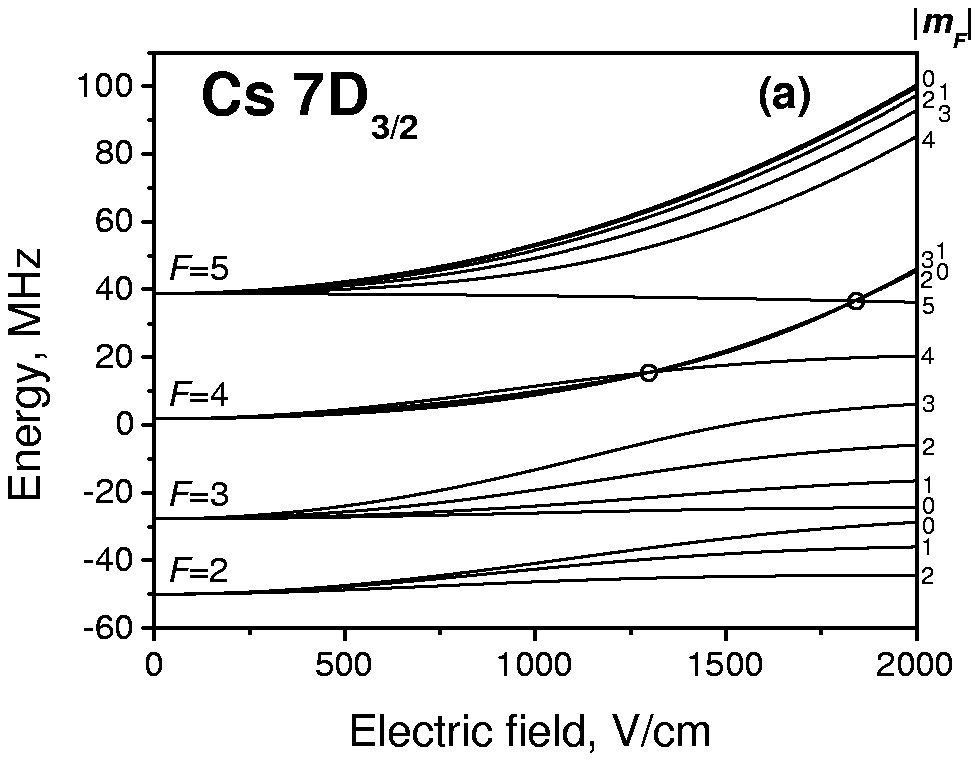}
\includegraphics[width=7.5cm]{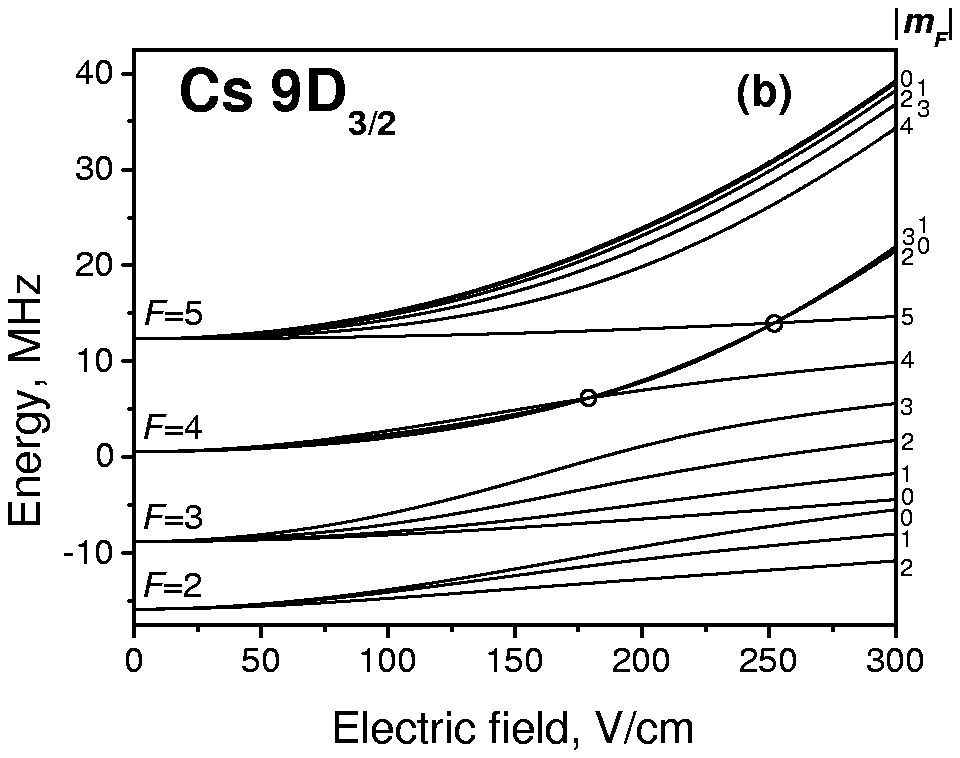}
\includegraphics[width=7.5cm]{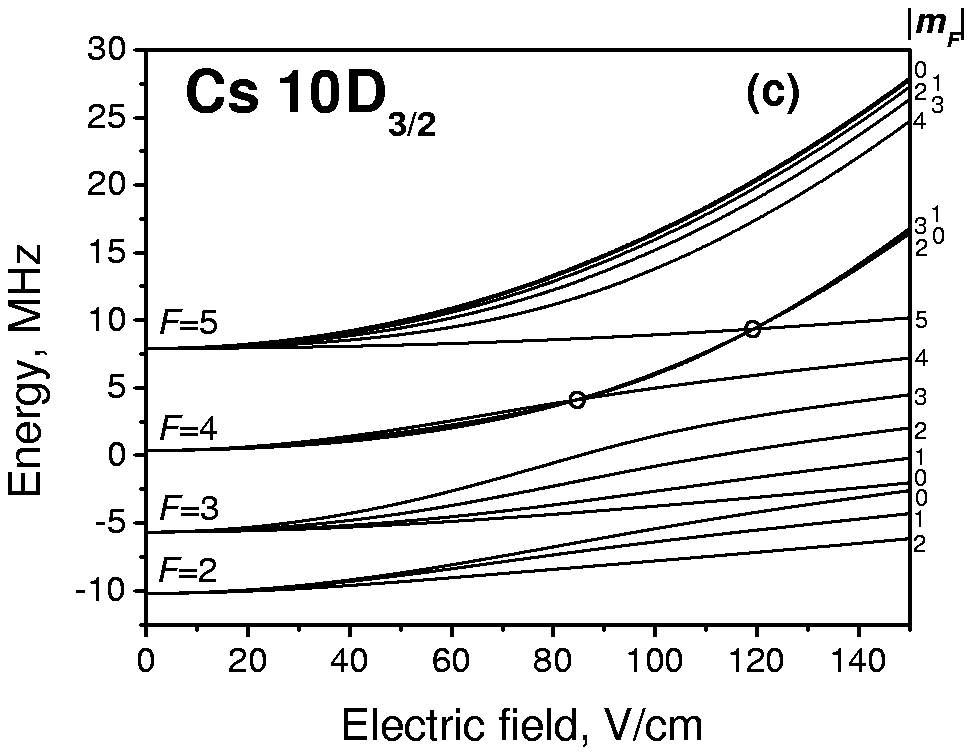}
\caption{Hyperfine level splitting diagram in an external electric field for the 7D$_{3/2}$ (a),
9D$_{3/2}$ (b), and 10D$_{3/2}$ (c) states of Cs, with zero energy corresponding to the fine
structure level energy.}
\label{f1}
\end{figure}

\begin{figure}
\begin{center}
\includegraphics[width=7.5cm]{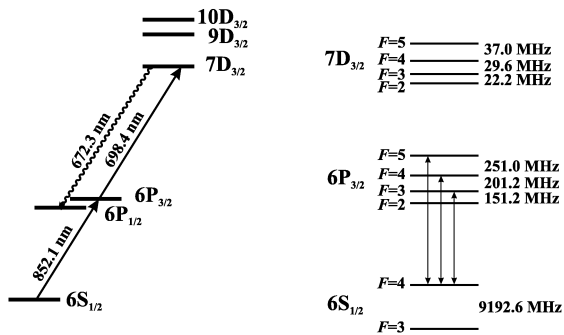}
\end{center}
\caption{Cesium energy-level scheme with hfs level spacings shown on the right
of the figure.}
\label{f2}
\end{figure}

In our experiment, we detected the resonance signals caused by hfs level-crossings in an
external dc electric field when several $m_F$ Zeeman sublevels of hfs levels were coherently
excited.  Figure \ref{f1} illustrates the crossing points of the hyperfine sublevels in
the (7,9,10)D$_{3/2}$ states.  The stepwise two-laser excitation
6S$_{1/2}$ $\rightarrow$ 6P$_{3/2} \rightarrow n$D$_{3/2}$ of the 7D$_{3/2}$,
$9$D$_{3/2}$ and $10$D$_{3/2}$
levels of atomic cesium was followed by the $n$D$_{3/2} \rightarrow$ 6P$_{1/2}$
fluorescence, as
shown in Fig. \ref{f2}.  The fluorescence intensity signal as a function of the electric
field strength is expected to contain resonances at positions corresponding to the
$m_F$ level-crossings. To predict the resonance positions, we computed
the energy level splitting  diagram in the presence of an electric field for hfs levels
in the  $n$D$_{3/2}$ states of Cs under study (see Fig. \ref{f1}) using
$\alpha_2$ values calculated by Wijngaarden and Li \cite{Wijngaarden94}.
We performed this calculation, whose results are shown in Fig. \ref{f1}
by diagonalizing the hfs and Stark interaction Hamiltonian
written in an uncoupled basis \cite{Aleksandrov93}.
The values for the hfs constant $A$
were taken from the review of Arimondo and collaborators
\cite{Arimondo77},
in which the following values were given: $A$ = 7.4(2) MHz for
7D$_{3/2}$ as measured in \cite{Belin76b},  $A$ = 2.35(4) MHz for 9D$_{3/2}$,  and
$A$ = 1.51(2) MHz for 10D$_{3/2}$ as measured in \cite{Deech75,Svanberg75}.
To our knowledge, no other experimental data of hfs constants for these states of Cs
were present in literature.
Values for the hfs constant $B$ were not reported and were assumed
to be small.

As seen in Fig. \ref{f1}, two crossings were predicted in the
experimentally available electric field range: one crossing within the
$F = 4$ manifold
with $\Delta m_F = \pm1$ and $\pm2$ and a second  $\Delta m_F = \pm2$ crossing between
the $m_F = \pm5$ sublevels of the $F = 5$ level and the $m_F = \pm 3$
sublevel of the $F = 4$ level. When the atoms were
excited with linearly polarized light, and linearly polarized fluorescence light was observed,
resonances were expected at the electric field values corresponding the the
level-crossings with
$\Delta m_F = \pm 2$ \cite{Alnis01,Auzinsh93}.

\subsection{Experimental details}
\begin{figure}
\begin{center}
\includegraphics[width=7.5cm]{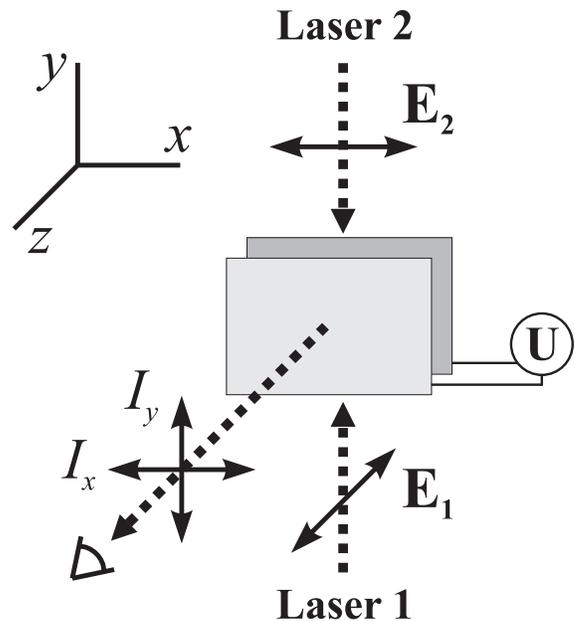}
\end{center}
\caption{Schematic diagram of the experiment.  The electrodes produced an electric field
along the $z$-axis, which was also parallel to the polarization vector of the first
laser {\bf E$_1$} and the observation direction.  The observed LIF intensity $I_x$ or $I_y$ polarization
direction could be chosen by linear polarizers.}
\label{f3}
\end{figure}

The schematic diagram of the experiment is depicted in Fig. \ref{f3}. In our experiment, cesium
vapor was produced in a sealed glass cell kept at room temperature. An electric field up
to $\mathbf{\mathcal{E}}$ = 2400 V/cm was applied via transparent Stark electrodes located inside the
cell.  Therefore, fluorescence light could be observed in the direction of the electric field.
The electrodes were separated by two ceramic spacer-rods, which had a diameter of 2.5 mm.
The transparent
electrodes were produced by depositing Indium-Tin-Oxide vapor on two glass plates. High
voltage could be applied to these electrodes using two metal rods which protruded through
the glass cell wall. High-temperature conducting silver paste was used to provide a
contact between the electrodes and the metal rods.

The 7D$_{3/2}$, 9D$_{3/2}$, and 10D$_{3/2}$ states of cesium were studied
using two-step laser excitation (see Fig. \ref{f2}). For the first step, 852.1 nm
radiation of the diode laser (LD-0850-100sm laser diode) was used to excite the 6P$_{3/2}$
state. The first laser was linearly polarized with polarization vector {\bf E$_1$}
along the external dc electric field
$\mathbf{\mathcal{E}}$ direction ($\mathbf{\mathcal{E}} \parallel$ {\bf z}). Radiation from
the second laser,
polarized as {\bf E$_2$} $\parallel$ {\bf x}, was sent in a counter-propagating direction to
induce either the 6P$_{3/2} \rightarrow $7D$_{3/2}$ transition at 698.4 nm using a
Hitachi HL6738MG laser diode or the 6P$_{3/2} \rightarrow$ 9D$_{3/2}$ transition at
584.8 nm and the 6P$_{3/2} \rightarrow$ 10D$_{3/2}$ transition at 563.7 nm using a
Coherent CR699-21 ring dye laser with Rodamin 6G dye. The dye laser was pumped by a Spectra-Physics 171
argon ion laser operating at the 514.5 nm line. The laser induced fluorescence (LIF)
$n$D$_{3/2} \rightarrow$ 6P$_{1/2}$ was observed at 672.3 nm, 566.4 nm, and 546.6 nm,
for $n=7,9$, and 10, respectively.  Before being observed, the LIF
passed through a linear polarizer. The observation direction was along
the $z$-axis.  Hence, we could observe the LIF intensity components
$I_x$ and $I_y$ (see Fig. \ref{f3}), which
were polarized parallel and perpendicular to {\bf E$_2$}, respectively.

In order to excite the cesium atoms from the ground state hyperfine level with total
angular momentum quantum number $F=4$ to all allowed 6P$_{3/2}$ state hyperfine levels
$F=3,4,5$, the first laser was operated in a multi-mode regime.  When the second laser
was the diode laser, we applied a 10-20 Hz saw-tooth signal to the piezo-electric crystal
mounted to its grating in order to jitter its output frequency over a range of 1.2 GHz.
The dye laser was operated in a single-mode regime.  To avoid
optical pumping, neutral density filters were used to reduce the dye laser intensity.
The power of the diode and dye lasers were not more than 3 and 10 mW, respectively.
The laser beams had a diameter of approximately 1 mm.
The frequency of both lasers was adjusted to maximize the observed fluorescence intensities at the
beginning of each measurement.

The LIF was focused by means of a two-lens system onto the entrance slit of a model MDR-3
monochromator with 1.3 nm/mm inverse dispersion and detected with a model FEU-79
photomultiplier tube, which was operated in photon counting mode.
The intensities $I_x$ and $I_y$ of
the LIF were detected as a function of
$\mathbf{\mathcal{E}}$. During the experiment, the high voltage between the electrodes was
scanned continuously. The photon counts were accumulated during one second intervals and
recorded on a PC together with the electrode voltage using a voltage divider. Signals
were accumulated during more than 10 scans of approximately 100 s duration,
binned and averaged.  We could adjust
for the delay between the voltage that was recorded and the actual voltage during the
integration interval by comparing the results recorded with rising and falling voltages.

\subsection{Results}
\begin{figure}
\begin{center}
\includegraphics[width=7cm]{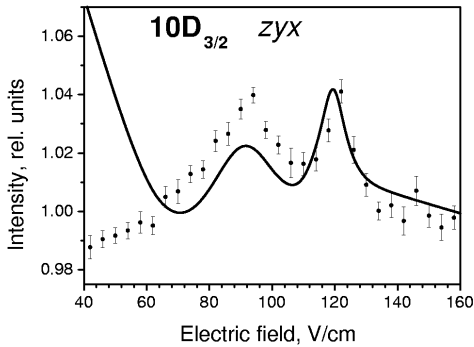}
\end{center}
\caption{Fluorescence vs. electric field for the 10D$_{3/2}$ state, $zyx$ geometry.
Dots, measurement; solid line, calculation.
The voltage scale before calibration was plotted.}
\label{f4}
\end{figure}

\begin{figure}
\includegraphics[width=7cm]{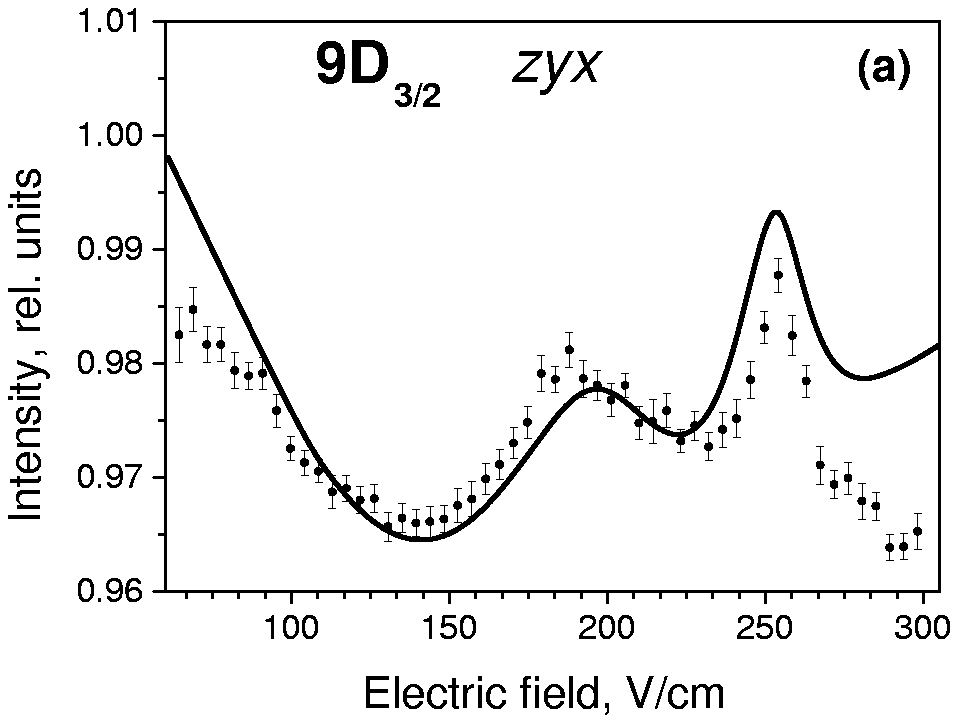}
\includegraphics[width=7cm]{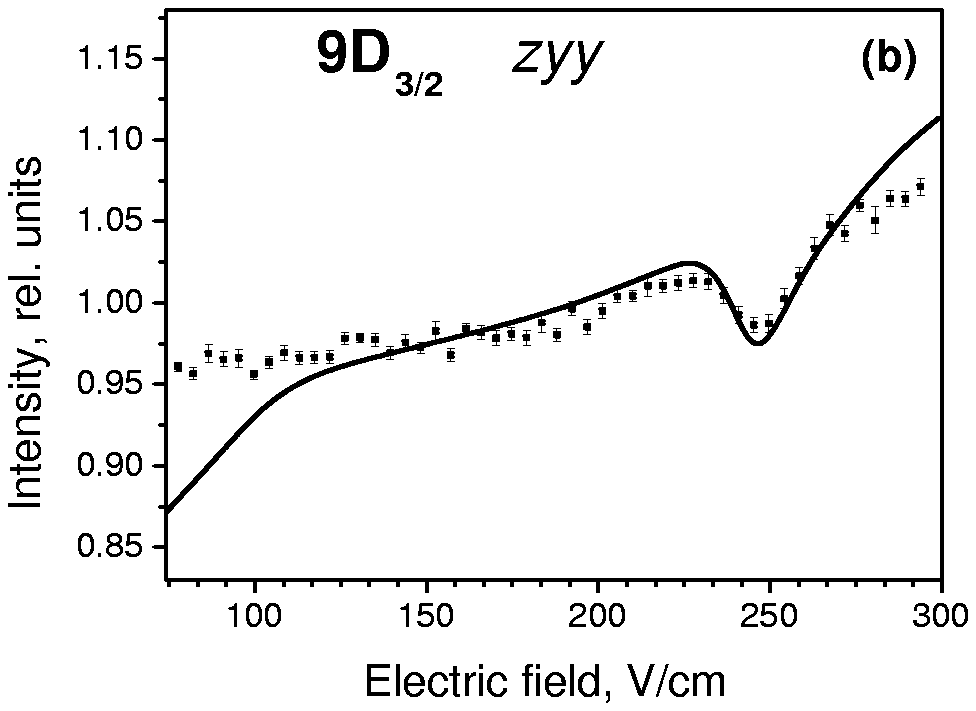}
\caption{Fluorescence vs. electric field for the 9D$_{3/2}$ state, $zyx$ geometry (a) and $zyy$ (b).
Dots, measurements; solid line, calculation.
The voltage scale was calibrated using data from Fig. \ref{f4}.}
\label{f5}
\end{figure}

\begin{figure}
\includegraphics[width=7cm]{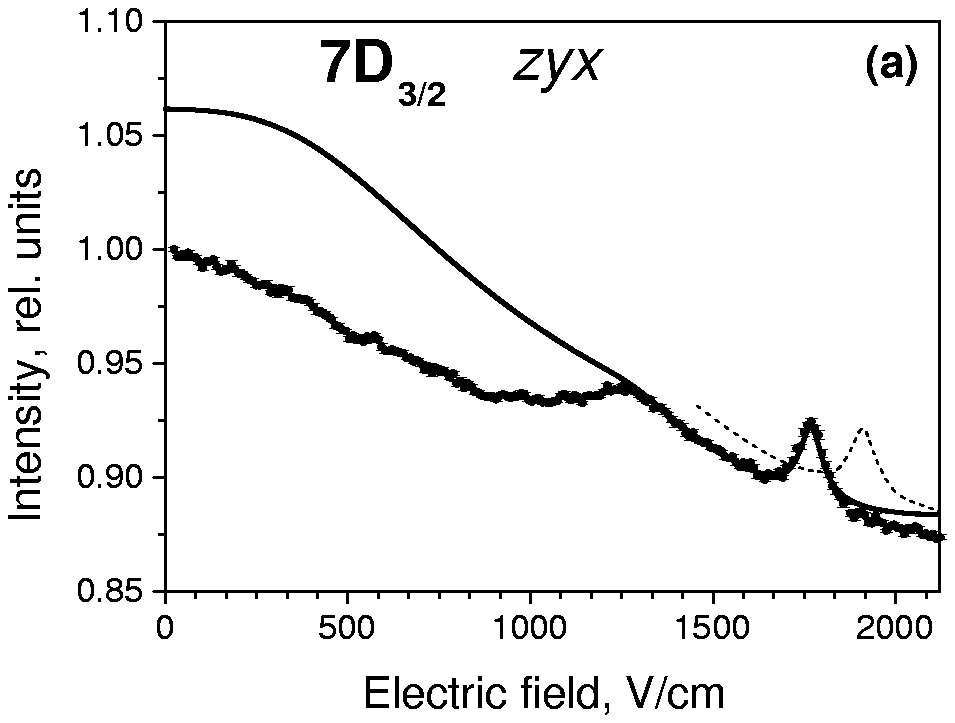}
\includegraphics[width=7cm]{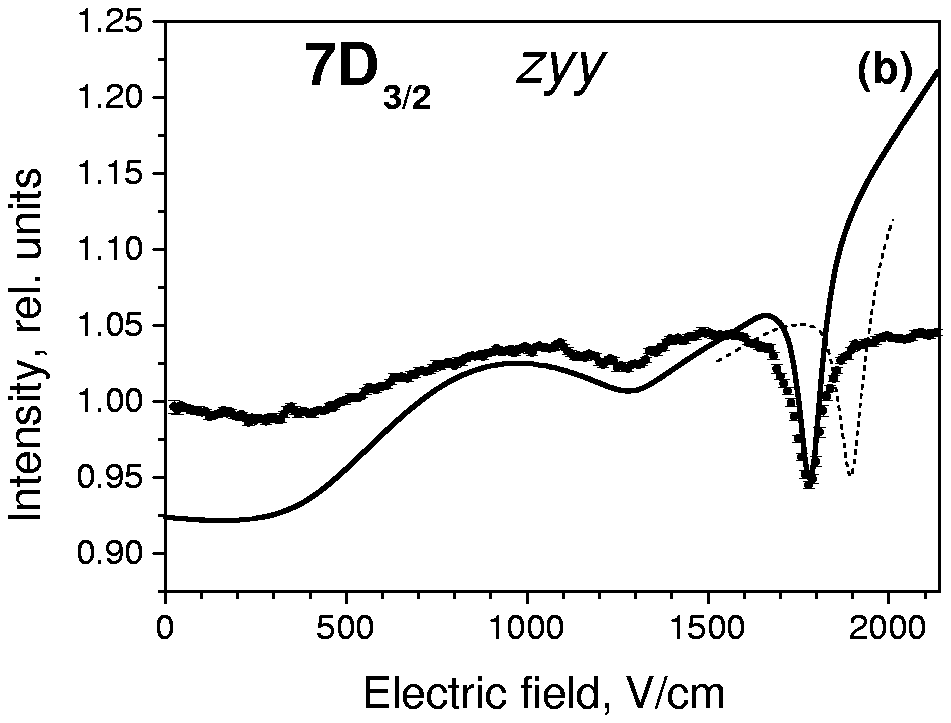}
\caption{Fluorescence vs. electric field for the 7D$_{3/2}$ state, $zyx$ geometry (a) and $zyy$ (b).
Dots, measurements; solid line, calculation; dashed line, calculation using the tensor
polarizability value from \cite{Wessel87}.
The voltage scale was calibrated using data from Fig. \ref{f4}.}
\label{f6}
\end{figure}

The measured relative fluorescence intensity was plotted as a function of electric field
strength in Figs. \ref{f4}--\ref{f6}.  The measured signals were represented by
dots, whereas solid lines were plotted to represent the results of
simulations.
The model on which these simulations were based was described
in Section 3 of this paper.  The error bars reflected the statistical variation in each bin
after the scans were averaged.  We labeled
the experimental geometry as $zyx$ or $zyy$: the first letter $z$ denoted
the orientation of the polarization
of the first laser {\bf E$_1$} (see Fig. \ref{f3}), the second letter $y$
denoted the orientation of {\bf E$_2$},
and the third letter $x$ or $y$ denoted the direction of LIF polarization that we observed.

The measurements for the 10D$_{3/2}$ state were plotted in Fig. \ref{f4}.
Since the tensor polarizability $\alpha_2$ for the 10D$_{3/2}$ state was known far better
than for the other states \cite{Xia97}, we used the data in Fig. \ref{f4} to calibrate the voltage.
In Fig. \ref{f4}, the voltage scale was left uncalibrated to illustrate the precision
with which the electrode spacing was known before calibration.
To simulate the results, we used
the hyperfine constant $A$ and experimentally determined
tensor polarizability $\alpha_2$ shown in Table \ref{t1}.
By comparing the position
of the second peak (corresponding to the $F$=4 to $F$=5 crossing) in our measured curve with
the peak position of the calculated curve, we determined that the
voltage scale had to be corrected by 2\%.

The results for the 9D$_{3/2}$ and 7D$_{3/2}$ states were plotted in
Figs. \ref{f5} and \ref{f6}, respectively.
The voltage scales in Figs. \ref{f5} and \ref{f6} have been adjusted using the
scaling factor obtained from the calibration with the 10D$_{3/2}$ signal in Fig. \ref{f4}.
The solid lines in Figs. \ref{f5} and \ref{f6} represented the result of calculations,
which were performed using
the hyperfine constant $A$ from Table \ref{t1} and a tensor polarizability
$\alpha_2$ adjusted so that the peak positions in the simulations and measured
data matched.
To illustrate the sensitivity of our
method, we included in Fig. \ref{f6} as a dashed line the results of a calculation using the previously measured
$\alpha_2$ values
shown in Table \ref{t1}.

\section{Theoretical Model}
\subsection{Outline of the model}
In the experiment described above, atoms strongly interacted with
radiation simultaneously produced by two lasers.
Nonlinear interactions can
cause shifts of the magnetic levels in the laser field \cite{Hap70},
and as a
result, shifts of the level-crossing positions. The theoretical description
of our experiment was further complicated by the fact, that in order to excite
coherently magnetic sublevels that were split in an external field, we used
the lasers that generated a rather broad profile of radiation.
In this situation, a model that was able
to describe signals quantitatively was essential in order to analyze
the obtained signals and to be able to deduce atomic
constants from these signals.

In the present study, such a model was elaborated on the basis of an approach
that we developed for the rate equations for Zeeman coherences in the case where
atoms were excited by one partially coherent optical field \cite{Blush04}.

In this paper, we extended this approach to the case when an atom interacts
with two laser fields simultaneously in presence of an external dc
electric field $\mathbf{\mathcal{E}}$. We assumed that the atomic center of mass moved
classically, which meant that the description of the dipole interaction of the atom
with the laser fields could assume a classically moving atom that was excited at
the internal transitions.
In this case the internal atomic dynamics could
be described by the semiclassical atomic density matrix $\rho $,\ which
parametrically depended on the classical coordinates of the atomic center of
mass.

We considered the absorption of the first laser's radiation as the atoms
were excited from the atomic
ground state denoted by $g$ to the intermediate state denoted by $e$. Then a
second laser excited the atoms further from the $e$ state to the final state
$f$. The direct transition $g\leftrightarrow f$
was forbidden in the dipole approximation.
In our particular case (see Fig. \ref{f2}), the ground state of the Cs
atom consisted of two
hfs levels $F_{g}=3$ and $F_{g}=4$. Each of these hfs levels in turn
consisted of $2F_{g}+1$ magnetic sublevels, denoted by $g_{i}$ in what
follows. The intermediate state in
our experiment was the 6P$_{3/2}$ state of the Cs atom. It consisted of four
hyperfine levels with $F_{e}=2,3,4$ and $5$ and the corresponding number of
magnetic sublevels, denoted as $e_{i}$.
Finally, the atomic level that was excited by the second laser was the $n$D$_{3/2}$
state, which again consisted of hyperfine levels with $F_{f}=2,3,4$ and $5$.
We denoted the magnetic sublevels of these states as $f_{i}$.

To simulate the observed signals, we had to
take into account that the external electric field was strong enough to break
partially the hyperfine interaction between electronic angular momentum of
the atom and the nuclear spin. As a result (see Fig. \ref{f1})
the magnetic
sublevel energies in the external dc electric field did not depend
quadratically on the electric field strength any more. This dependence could
be obtained only by diagonalizing the full Hamilton matrix. The partial
decoupling of the electronic angular momentum and nuclear spin altered also
the dipole transition probabilities between the magnetic sublevels of the atoms
belonging to the different fine structure levels. This decoupling was taken into
account in the simulation of the experimental signals.

With the above assumptions, we were able to develop a model,
described below, to calculate the observed level-crossing signals in our
experiment.

\subsection{Optical Bloch equations}
We started our analysis from the optical Bloch equations (OBEs) for the density
matrix elements $\rho _{g_{i}g_{j}}$, $\rho _{g_{i}e_{j}}$, $\rho
_{g_{i}f_{j}}$, $\rho _{e_{i}g_{j}}$, $\rho _{e_{i}e_{j}}$, $\rho _{e_{i}f_{j}}$,
 $\rho_{f_{i}g_{j}}$, $\rho _{f_{i}e_{j}}$, and $\rho _{f_{i}f_{j}}$. In writing OBEs (see
for example \cite{Ste84}),
\begin{equation}
i\hbar \dfrac{\partial \rho }{\partial t}=\left[ \widehat{H},\widetilde{\rho
}\right] +i\hbar \widehat{R}\rho ,
\end{equation}%
we considered the relaxation $\widehat{R}$ operator to include spontaneous
emission and transit relaxation due to the thermal motion of atoms into and out
of the laser beam. We also assumed that different velocity groups of the thermally
moving atoms did not interact -- the density of atoms was sufficiently low. In
this case the relaxation matrix was:

\begin{center}
\begin{eqnarray}
\widehat{R}\rho_{g_{i}g_{j}}&=&\underset{e_{i}e_{j}}{\sum}\Gamma_{g_{i}g_{j}%
}^{e_{i}e_{j}}\rho_{e_{i}e_{j}}-\gamma\rho_{g_{i}g_{j}}+\lambda\delta_{g_{i},g_{j}},
\notag \\
\widehat{R}\rho_{g_{i}e_{j}}&=&-\dfrac{\Gamma_{e}}{2}\rho_{g_{i}e_{j}}%
-\gamma\rho_{g_{i}e_{j}}, \notag \\%
\widehat{R}\rho_{g_{i}f_{j}}&=&-\dfrac{\Gamma_{f}}{2}\rho_{g_{i}f_{j}}%
-\gamma\rho_{g_{i}f_{j}}, \notag \\%
\widehat{R}\rho_{e_{i}g_{j}}&=&-\dfrac{\Gamma_{e}}{2}\rho_{e_{i}g_{j}}%
-\gamma\rho_{e_{i}g_{j}}, \notag \\%
\widehat{R}\rho_{e_{i}e_{j}}&=&-\Gamma_{e}\rho_{e_{i}e_{j}}+\underset{f_{i}%
f_{j}}{\sum}\Gamma_{e_{i}e_{j}}^{f_{i}f_{j}}\rho_{f_{i}f_{j}}-\gamma
\rho_{e_{i}e_{j}}, \\%
\widehat{R}\rho_{e_{i}f_{j}}&=&-\left(  \dfrac{\Gamma_{e}}{2}+\dfrac{\Gamma_{f}%
}{2}\right)  \rho_{e_{i}f_{j}}-\gamma\rho_{e_{i}f_{j}}, \notag \\%
\widehat{R}\rho_{f_{i}g_{j}}&=&-\dfrac{\Gamma_{f}}{2}\rho_{f_{i}g_{j}}%
-\gamma\rho_{f_{i}g_{j}}, \notag \\%
\widehat{R}\rho_{f_{i}e_{j}}&=&-\left(  \dfrac{\Gamma_{e}}{2}+\dfrac{\Gamma_{f}%
}{2}\right)  \rho_{f_{i}e_{j}}-\gamma\rho_{f_{i}e_{j}}, \notag \\%
\widehat{R}\rho_{f_{i}f_{j}}&=&-\Gamma_{f}\rho_{f_{i}f_{j}}-\gamma\rho
_{f_{i}f_{j}}, \notag
\end{eqnarray}
\end{center}
where $\gamma $ and $\lambda $ were the transit relaxation rates. The quantity $\lambda
\delta _{g_{i},g_{j}}\ $described the process in which \textquotedblleft
fresh" atoms were moving into the laser beam, and $\gamma $ described the
rate at which atoms were leaving the interaction region. $\Gamma _{e}$ was the total
spontaneous relaxation rate from level $e$, $\Gamma _{f}$ was the total
spontaneous relaxation rate from level $f$, $\Gamma
_{g_{i}g_{j}}^{e_{i}e_{j}}$ described the spontaneous relaxation from $\rho
_{e_{i}e_{j}}$ to $\rho _{g_{i}g_{j}}$, $\Gamma _{e_{i}e_{j}}^{f_{i}f_{j}}$
described the spontaneous relaxation from $\rho _{f_{i}f_{j}}$ to $\rho
_{e_{i}e_{j}}$.  The explicit forms of these rate coefficients were calculated
on the basis of angular momentum algebra and can be found in \cite{Auz05}.

The Hamiltonian $\widehat{H}=\widehat{H}_{0}+\widehat{V}$ included the
unperturbed atomic Hamiltonian $\widehat{H}_{0}$ and the dipole interaction
operator $\widehat{V}=-\widehat{\mathbf{d}}\cdot \mathbf{E}\left( t\right) $%
, where $\widehat{\mathbf{d}}$\ was the electric dipole operator.\ The
exciting light was described classically by two uncorrelated fluctuating
electric fields $\mathbf{E_1}$ and $\mathbf{E_2}$
of definite polarizations $\mathbf{e}_{1}$ and $\mathbf{e}%
_{2}$:
\begin{center}
\begin{eqnarray}
\mathbf{E}\left( t\right) &=&\mathbf{E}_{1}\left( t\right) +\mathbf{E}%
_{2}\left( t\right),  \notag \\
\mathbf{E}_{i}\left( t\right) &=&\varepsilon _{i}\left( t\right) \mathbf{e}%
_{i}+\varepsilon _{i}^{\ast }\left( t\right) \mathbf{e}_{i}^{\ast }, \\
\varepsilon _{i}(t) &=&\left\vert \varepsilon _{\overline{\omega}_{i}%
}\right\vert \exp \left[ -i\Phi _{i}\left( t\right) -i\left( \overline{%
\omega}_{i} \mp \mathbf{k}_{\overline{\omega}_{i}}\mathbf{v}\right) t\right],
\notag
\end{eqnarray}
\end{center}

with the center frequency of the radiation spectrum $\overline{\omega}_{i}$\ and the
fluctuating phase $\Phi _{i}\left( t\right) $.  The lineshape of the exciting light
was assumed to be Lorentzian with FWHM $\Delta \omega _{i}$. Atoms moved with
definite velocity $\mathbf{v}$, which gave the shift $\overline{\omega}_{i}%
\mp \mathbf{k}_{\overline{\omega}_{i}}\mathbf{v}$ in the laser frequency that
the atom would encounter due to the Doppler effect,
where $\mathbf{k}_{\overline{\omega} _{i}}$ was the wave vector of the
exciting light.  The minus sign referred to the laser beam that propagated in the
positive direction of the $y$-axis (see Fig. \ref{f3}) and the plus sign to the counterpropagating
laser beam.

Writing OBEs explicitly for the density matrix elements $\rho_{ij}$, we obtained:

\begin{widetext}

\begin{center}
\begin{eqnarray}
\dfrac{\partial \rho _{ij}}{\partial t} &=&-\dfrac{i}{\hbar }\left[ \widehat{%
H},\rho _{ij}\right] +\widehat{R}\rho _{ij}=  \notag \\
&=&-\dfrac{i}{\hbar }\left[ \widehat{H}_{0},\rho _{ij}\right] +\dfrac{i}{%
\hbar }\left[ \widehat{\mathbf{d}}\cdot \mathbf{E}\left( t\right) ,\rho _{ij}%
\right] +\widehat{R}\rho _{ij}= \\
&=&-i\omega _{ij}\rho _{ij}+\dfrac{i}{\hbar }\mathbf{E}\left( t\right)
\underset{k}{\sum }\left( \mathbf{d}_{ik}\cdot \rho _{kj}-\rho _{ik}\cdot
\mathbf{d}_{kj}\right) +\widehat{R}\rho _{ij},  \notag
\end{eqnarray}
\end{center}

\end{widetext}

where $\omega _{ij}=\omega_{i}-\omega_{j}$ denoted the splitting of
the levels $i$ and $j$ and $\mathbf{d}_{ik}\equiv \left\langle i\left\vert
\mathbf{d}\right\vert k\right\rangle $. By choosing the quantization axis (the $z$-axis) to be
parallel to the static electric field $\mathbf{\mathcal{E}}$, all the explicit dependence of the density
matrix on the static electric field $\mathbf{\mathcal{E}}$ was included in the splitting terms $\omega _{ij}$%
. Implicitly, the density matrix depended on the dc electric field, because
this field modified the dipole transition matrix elements by
partially decoupling the hyperfine interaction.

In order to simplify the above equation, we did the following: we neglected
possible optical excitations of neighboring transitions, that is, we
neglected the excitation of the transitions $g\leftrightarrow e$ ($%
e\leftrightarrow f$) with the second (first) laser, which was tuned to the
transition $e\leftrightarrow f$ ($g\leftrightarrow e$). Then, in order to
eliminate fast oscillations with optical frequencies $\overline{\omega} _{i}$%
, we applied to the optical Bloch equations the rotating wave approximation
for multilevel systems as developed in \cite{Ari96}:

\begin{widetext}

\begin{center}
\begin{eqnarray}
\rho _{ge} &=&\widetilde{\rho}_{ge}e^{i\left( \overline{\omega}_{1}-%
\mathbf{k}_{\overline{\omega}_{1}}\boldsymbol{v}\right) t+i\Phi _{1}\left(
t\right) }=\rho _{eg}^{\ast },  \notag \\
\rho _{gf} &=&\widetilde{\rho} _{gf}e^{i\left( \overline{\omega} _{1}+%
\overline{\omega} _{2}-\mathbf{k}_{\overline{\omega} _{1}}\boldsymbol{v}+%
\mathbf{k}_{\overline{\omega} _{2}}\boldsymbol{v}\right) t+i\Phi _{1}\left(
t\right) +i\Phi _{2}\left( t\right) }=\rho _{fg}^{\ast }, \\
\rho _{ef} &=&\widetilde{\rho} _{ef}e^{i\left( \overline{\omega} _{2}+%
\mathbf{k}_{\overline{\omega} _{2}}\boldsymbol{v}\right) t+i\Phi _{2}\left(
t\right) }=\rho _{fe}^{\ast }.  \notag
\end{eqnarray}
\end{center}

\end{widetext}

\subsection{Laser radiation fluctuations}

In the optical Bloch equations we distinguished Zeeman coherences that
corresponded to the density matrix elements $\widetilde{\rho} _{gg}$, $\widetilde{%
\rho} _{ee}$, $\widetilde{\rho} _{ff}$ and optical coherences that corresponded
to the density matrix elements $\widetilde{\rho} _{ef}$, $\widetilde{\rho} _{fe}%
$, $\widetilde{\rho} _{ge}$, $\widetilde{\rho} _{eg}$.
As a result we arrived at a system of stochastic differential equations (4)
with stochastic variables $\Phi _{i}\left( t\right) $.  We simplified this
system by applying the \textquotedblleft decorrelation approach" \cite{Kam76}.

In the experiment we observed
signals that were averaged over time intervals that were large in
comparison with the characteristic phase-fluctuation time of the
excitation-light source.  Therefore we needed to perform a statistical
averaging of the above equations. In order to do that, we solved the
equations for optical coherences and then took a formal statistical average
over the fluctuating phases (for details see \cite{Blush04}). Additionally we
assumed that both lasers were uncorrelated and that optical coherences $%
\widetilde{\rho} _{ef},\widetilde{\rho} _{fe}$ ($\widetilde{\rho} _{ge},%
\widetilde{\rho} _{eg}$) were independent of the fluctuations of the first
(second) laser, which was tuned to the transition $g\leftrightarrow e$ ($%
e\leftrightarrow f$). Then we applied the \textquotedblleft decorrelation
approximation" (see \cite{Blush04} and references cited therein):

\begin{center}
\begin{equation}
\left\langle \rho _{ij}\left( t^{\prime }\right) e^{\pm i\left[ \Phi \left(
t\right) -\Phi \left( t^{\prime }\right) \right] }\right\rangle
=\left\langle \rho _{ij}\left( t^{\prime }\right) \right\rangle \left\langle
e^{\pm i\left[ \Phi \left( t\right) -\Phi \left( t^{\prime }\right) \right]
}\right\rangle.
\end{equation}
\end{center}

The correlation function $\left\langle e^{\pm i\left[ \Phi \left( t\right)
-\Phi \left( t^{\prime }\right) \right] }\right\rangle $ was calculated
assuming the \textquotedblleft phase diffusion" model of the laser radiation for
the description of the dynamics of the fluctuating phase \cite{Blush04}. Thus,

\begin{center}
\begin{equation}
\left\langle e^{\pm i\left[ \Phi (t)-\Phi (t^{\prime })\right]
}\right\rangle =e^{-\frac{\triangle \omega }{2}\left( t-t^{\prime }\right) }.
\end{equation}
\end{center}

Putting it all together, we arrived at the phase-averaged OBEs (for simplicity
we dropped the averaging brackets). In the case of stationary time-independent
excitation we obtained

\begin{widetext}

\begin{center}
\begin{eqnarray}
\rho _{g_{i}g_{j}} &=&\dfrac{i}{\hbar }\dfrac{\left\vert \varepsilon _{%
\overline{\omega} _{1}}\right\vert }{\gamma +i\omega _{gigj}}\underset{e_{k}}%
{\sum }\left( d_{g_{i}e_{k}}^{\left( 1\right) \ast }\widetilde{\rho}
_{e_{k}g_{j}}-d_{e_{k}g_{j}}^{\left( 1\right) }\widetilde{\rho} _{g_{i}e_{k}%
}\right) + \notag \\
&&+\dfrac{1}{\gamma +i\omega _{gigj}}\left( \underset{e_{i}e_{j}}{\sum }%
\Gamma _{g_{i}g_{j}}^{e_{i}e_{j}}\rho _{e_{i}e_{j}}+\lambda \delta \left(
g_{i},g_{j}\right) \right) ,  \notag
\end{eqnarray}

\begin{eqnarray}
\widetilde{\rho} _{g_{i}e_{j}} &=&\dfrac{i}{\hbar }\dfrac{1}{\left( \frac{%
\Gamma _{e}}{2}+\gamma +\frac{\triangle \omega _{1}}{2}\right) +i\left(
\overline{\omega} _{1}-\mathbf{k}_{\overline{\omega} _{1}}\boldsymbol{v}%
+\omega _{giej}\right) }\times  \notag \\
&&\times \left( \left\vert \varepsilon _{\overline{\omega} _{1}}\right\vert
\underset{e_{k}}{\sum }d_{g_{i}e_{k}}^{\left( 1\right) \ast }\rho
_{e_{k}e_{j}}-\left\vert \varepsilon _{\overline{\omega} _{1}}\right\vert
\underset{g_{k}}{\sum }d_{g_{k}e_{j}}^{\left( 1\right) \ast }\rho
_{g_{i}g_{k}}-\left\vert \varepsilon _{\overline{\omega} _{2}}\right\vert
\underset{f_{k}}{\sum }d_{f_{k}e_{j}}^{\left( 2\right) }\widetilde{\rho}
_{g_{i}f_{k}}\right) ,  \notag
\end{eqnarray}

\begin{eqnarray}
\widetilde{\rho} _{g_{i}f_{j}} &=&\dfrac{i}{\hbar }\dfrac{1}{\left( \frac{%
\Gamma _{f}}{2}+\gamma +\frac{\triangle \omega _{1}}{2}+\frac{\triangle
\omega _{2}}{2}\right) +i\left( \overline{\omega} _{1}+\overline{\omega} _{2}%
-\mathbf{k}_{\overline{\omega} _{1}}\boldsymbol{v}+\mathbf{k}_{\overline{%
\omega} _2}\boldsymbol{v}+\omega _{g_{i}f_{j}}\right) }\times  \notag \\
&&\times \underset{e_{k}}{\sum }\left( \left\vert \varepsilon _{\overline{%
\omega} _{1}}\right\vert d_{g_{i}e_{k}}^{\left( 1\right) \ast }\widetilde{%
\rho} _{e_{k}f_{j}}-\left\vert \varepsilon _{\overline{\omega} _{2}%
}\right\vert d_{e_{k}f_{j}}^{\left( 2\right) \ast }\widetilde{\rho}
_{g_{i}e_{k}}\right) ,  \notag
\end{eqnarray}

\begin{eqnarray}
\widetilde{\rho} _{e_{i}g_{j}} &=&\dfrac{i}{\hbar }\dfrac{1}{\left( \frac{%
\Gamma _{e}}{2}+\gamma +\frac{\triangle \omega _{1}}{2}\right) -i\left(
\overline{\omega} _{1}-\mathbf{k}_{\overline{\omega} _{1}}\boldsymbol{v}%
-\omega _{e_{i}g_{j}}\right) }\times  \notag \\
&&\times \left( \left\vert \varepsilon _{\overline{\omega} _{1}}\right\vert
\underset{g_{k}}{\sum }d_{e_{i}g_{k}}^{\left( 1\right) }\rho
_{g_{k}g_{j}}-\left\vert \varepsilon _{\overline{\omega} _{1}}\right\vert
\underset{e_{k}}{\sum }d_{e_{k}g_{j}}^{\left( 1\right) }\rho
_{e_{i}e_{k}}+\left\vert \varepsilon _{\overline{\omega} _{2}}\right\vert
\underset{f_{k}}{\sum }d_{e_{i}f_{k}}^{\left( 2\right) \ast }\widetilde{\rho}
_{f_{k}g_{j}}\right) ,  \notag
\end{eqnarray}

\begin{eqnarray}
\rho _{e_{i}e_{j}} &=&\dfrac{i}{\hbar }\dfrac{\left\vert \varepsilon _{%
\overline{\omega} _{1}}\right\vert }{\left( \Gamma _{e}+\gamma \right)
+i\omega _{e_{i}e_{j}}}\underset{g_{k}}{\sum }\left( d_{e_{i}g_{k}}^{\left(
1\right) }\widetilde{\rho} _{g_{k}e_{j}}-d_{g_{k}e_{j}}^{\left( 1\right)
\ast }\widetilde{\rho} _{e_{i}g_{k}}\right) +  \\
&&+\dfrac{i}{\hbar }\dfrac{\left\vert \varepsilon _{\overline{\omega} _{2}%
}\right\vert }{\left( \Gamma _{e}+\gamma \right) +i\omega _{e_{i}e_{j}}}%
\underset{f_{k}}{\sum }\left( d_{e_{i}f_{k}}^{\left( 2\right) \ast }%
\widetilde{\rho} _{f_{k}e_{j}}-d_{f_{k}e_{j}}^{\left( 2\right) }\widetilde{%
\rho} _{e_{i}f_{k}}\right) +  \notag \\
&&+\dfrac{1}{\left( \Gamma _{e}+\gamma \right) +i\omega _{e_{i}e_{j}}}%
\underset{f_{i}f_{j}}{\sum }\Gamma _{e_{i}e_{j}}^{f_{i}f_{j}}\rho
_{f_{i}f_{j}},  \notag
\end{eqnarray}

\begin{eqnarray}
\widetilde{\rho} _{e_{i}f_{j}} &=&\dfrac{i}{\hbar }\dfrac{1}{\left( \frac{%
\Gamma _{e}}{2}+\frac{\Gamma _{f}}{2}+\gamma +\frac{\triangle \omega _{2}}{2}%
\right) +i\left( \overline{\omega} _{2}-\mathbf{k}_{\overline{\omega}_2}%
\boldsymbol{v}+\omega _{e_{i}f_{j}}\right) }\times  \notag \\
&&\times \left( \left\vert \varepsilon _{\overline{\omega} _{2}}\right\vert
\underset{f_{k}}{\sum }d_{e_{i}f_{k}}^{\left( 2\right) \ast }\rho
_{f_{k}f_{j}}-\left\vert \varepsilon _{\overline{\omega} _{2}}\right\vert
\underset{e_{k}}{\sum }d_{e_{k}f_{j}}^{\left( 2\right) \ast }\rho
_{e_{i}e_{k}}+\left\vert \varepsilon _{\overline{\omega} _{1}}\right\vert
\underset{g_{k}}{\sum }d_{e_{i}g_{k}}^{\left( 1\right) }\widetilde{\rho}
_{g_{k}f_{j}}\right) ,  \notag
\end{eqnarray}

\begin{eqnarray}
\widetilde{\rho} _{f_{i}g_{j}} &=&\dfrac{i}{\hbar }\dfrac{1}{\left( \frac{%
\Gamma _{f}}{2}+\gamma +\frac{\triangle \omega _{1}}{2}+\frac{\triangle
\omega _{2}}{2}\right) -i\left( \overline{\omega} _{1}+\overline{\omega} _{2}%
-\overrightarrow{k_{\overline{\omega} _{1}}}\overrightarrow{v}-%
\overrightarrow{k_{\overline{\omega} _{2}}}\overrightarrow{v}-\omega
_{f_{i}g_{j}}\right) }\times  \notag \\
&&\times \underset{e_{k}}{\sum }\left( \left\vert \varepsilon _{\overline{%
\omega _{2}}}\right\vert d_{f_{i}e_{k}}^{\left( 2\right) }\widetilde{\rho}
_{e_{k}g_{j}}-\left\vert \varepsilon _{\overline{\omega} _{1}}\right\vert
d_{e_{k}g_{j}}^{\left( 1\right) }\widetilde{\rho} _{f_{i}e_{k}}\right) ,
\notag
\end{eqnarray}

\begin{eqnarray}
\widetilde{\rho} _{f_{i}e_{j}} &=&\dfrac{i}{\hbar }\dfrac{1}{\left( \frac{%
\Gamma _{e}}{2}+\frac{\Gamma _{f}}{2}+\gamma +\frac{\triangle \omega _{2}}{2}%
\right) -i\left( \overline{\omega} _{2}-\mathbf{k}_{\overline{\omega} _{2}}%
\boldsymbol{v}-\omega _{f_{i}e_{j}}\right) }\times  \notag \\
&&\times \left( \left\vert \varepsilon _{\overline{\omega} _{2}}\right\vert
\underset{e_{k}}{\sum }d_{f_{i}e_{k}}^{\left( 2\right) }\rho
_{e_{k}e_{j}}-\left\vert \varepsilon _{\overline{\omega} _{2}}\right\vert
\underset{f_{k}}{\sum }d_{f_{k}e_{j}}^{\left( 2\right) }\rho
_{f_{i}f_{k}}-\left\vert \varepsilon _{\overline{\omega} _{1}}\right\vert
\underset{g_{k}}{\sum }d_{g_{k}e_{j}}^{\left( 1\right) \ast }\widetilde{\rho}
_{f_{i}g_{k}}\right) ,  \notag
\end{eqnarray}

\begin{equation}
\rho _{f_{i}f_{j}}=\dfrac{i}{\hbar }\dfrac{\left\vert \varepsilon _{%
\overline{\omega} _{2}}\right\vert }{\left( \Gamma _{f}+\gamma \right)
+i\omega _{f_{i}f_{j}}}\underset{e_{k}}{\sum }\left( d_{f_{i}e_{k}}^{\left(
2\right) }\widetilde{\rho} _{e_{k}f_{j}}-d_{e_{k}f_{j}}^{\left( 2\right)
\ast }\widetilde{\rho} _{f_{i}e_{k}}\right) . \notag
\end{equation}
\end{center}

\end{widetext}

This system of equations was the one that we solved in order to
simulate the observed signals.  When the density matrix for the final state
was calculated, we obtained the fluorescence intensities with a specific
polarization along the unit vector $\mathbf{e}$\ as \cite{Auz05,Coh61,Dya64}:%
\begin{equation}
I\left( \mathbf{e}\right) =\widetilde{I}_{0}\underset{g_{i},f_{i},f_{j}}{%
\sum }d_{g_{i}f_{j}}^{(ob)\ast }d_{e_{i}g_{i}}^{(ob)}\rho _{f_{i}f_{j}},
\label{Eq7.1}
\end{equation}%
where $\widetilde{I_{0}}$ was a proportionality coefficient and the matrix
element $d_{g_{i}f_{j}}^{(ob)}=$ $\left\langle g_{i}\left\vert \mathbf{d}%
\cdot \mathbf{e}\right\vert f_{j}\right\rangle $ contained the polarization
vector $\mathbf{e}$ of the light which was detected, i.e. along the
$x$- or $y$-axis.

\section{Analysis and discussion}
The theoretical model discussed above was used to simulate our experiment,
and the results of the simulations were plotted together
with the results of our measurements in Fig. \ref{f4}--\ref{f6} as described
in Section 2.
Since the precise shape of the level-crossing signal depended on various parameters
that were beyond our ability to control precisely, these parameters were adjusted in the
calculation.  The parameters that we adjusted were the Rabi frequencies of the transitions, the laser
radiation spectral widths, and the detuning of the laser radiation
relative to the exact transition frequencies.
In addition, the background was left as an adjustable parameter.
The generally good agreement between the calculation and the measurement, except at
electric fields far below and above the level-crossing points, validated the theoretical
approach described in Section 3.

The positions of the resonances depended on the points where energy levels crossed (see
Fig. \ref{f1}).  These, in turn, depended on the values of the hyperfine constants $A$
and $B$, and on the tensor polarizability $\alpha_2$.  We took the hyperfine constants
to be sufficiently well determined (see the review by Arimondo and collaborators \cite{Arimondo77})
to allow us to
use our results to make a new measurement of the tensor polarizabilities of the 9D$_{3/2}$
and 7D$_{3/2}$ states of cesium.  Our results were summarized in Table \ref{t1} and
compared with the previous measurements of Fredrikson and Svanberg \cite{Fredrikson77} and
of Wessel and Cooper \cite{Wessel87}, and with the theoretical calculations of $\alpha_2$
of Wijngaarden and Li \cite{Wijngaarden94}.  We estimated the accuracy of our value for
$\alpha_2$ based on the reproducibility
of several measurements and accounted for the uncertainty of the
hyperfine constant $A$ and of the tensor polarizability $\alpha_2$ of the 10D$_{3/2}$ state,
on which our electric field calibration was based.  Furthermore, we included the error
in the measurements of the polarizabilities of the 7D$_{3/2}$ and 9D$_{3/2}$
states introduced by the reported uncertainties in their respective hyperfine constants,
$A$.  The largest contributions to our error were the uncertainty in the hyperfine
constants and in the electric field calibration, whose contributions were of comparable magnitude.

Our accuracy was competitive or slightly higher than those of
previously reported measurements of the tensor polarizabilities for these atomic states.
Our results were consistent with the theoretical predictions for $\alpha_2$ of
\cite{Wijngaarden94}
for the 9D$_{3/2}$ state as well as with the previous measurement of \cite{Fredrikson77}.  For the 7D$_{3/2}$ state, our measurements indicated
a value for $\alpha_2$ that was higher than both the previous measurement of Wessel and
Cooper \cite{Wessel87} and the
theoretical prediction of \cite{Wijngaarden94}.

\begin{table*}
\caption{Summary of results.}
\begin{small}
\begin{tabular*}{\textwidth}{@{\extracolsep{\fill}}ccccc}
\hline
                &                                      &  \multicolumn{3}{c}{Tensor Polarizability $\alpha_2$}                                    \\ \cline{3-5}
Cesium          & Hyperfine                            &  This               & Previous                             & Theory                      \\
Atomic          & Constant                             &  experiment         & experiment                           &                             \\
State           & (MHz)                                &  ($a_0^3$)          & ($a_0^3$)                            & ($a_0^3)$                   \\
\hline
\hline
10D$_{3/2}$     & 1.51(2) \cite{Arimondo77}    &  --                  & $3.4012(36) \times 10^6$ \cite{Xia97}     & $3.41 \times 10^6$ \cite{Wijngaarden94} \\
9D$_{3/2}$      & 2.35(4) \cite{Arimondo77}    &  $1.183(35) \times 10^6$ & $1.258(60) \times 10^6$  \cite{Fredrikson77}  & $1.19 \times 10^6$ \cite{Wijngaarden94}\\
7D$_{3/2}$      & 7.4(2)  \cite{Arimondo77}              &  $7.45(20) \times 10^4$  & $6.6(3) \times 10^4$     \cite{Wessel87}  & $7.04 \times 10^4$ \cite{Wijngaarden94} \\
\hline
\end{tabular*}
\label{t1}
\end{small}
\end{table*}

\section{Concluding Remarks}
The method of detecting pure electric field induced level-crossing signals of $m_F$ Zeeman
sublevels of the hyperfine $F$ levels at two-step laser excitation has
been applied to determine experimentally the tensor
polarizabilities of highly excited atomic states. Conventional laser sources,
including diode lasers, with rather broad line contours were sufficient. In the
case of crossings between different $F$ sublevels with $\Delta m_F = \pm 2$,
the resonance peaks were sufficiently sharp to
enable accurate determination of the peak position. At the same time, the
fluorescence intensity behavior within a broader electric field range, including
additional crossings, together with reliable signal simulations, enhanced
the accuracy of the technique.

For this purpose an adequate theoretical description was worked out by
extending an approach previously developed for two-level systems
\cite{Blush04} to the case of a three-level system.  A significant
simplification of the optical Bloch equations was achieved by statistically
averaging over the fluctuating phases and applying the
\textquotedblleft decorrelation approximation”.  Though the problem was more cumbersome
with more parameters to be considered, it made it possible to describe
satisfactorily the observed signals.
What is more, moderate computation times could be achieved by
replacing the Doppler distribution with a group of atoms moving at a definite
velocity.

The measured tensor polarizability for the higher-$n$ 9D$_{3/2}$ state
(see Table \ref{t1}) agreed within experimental error with previously
measured and calculated values. At the same time the present measured
tensor polarizability for the lower-$n$ 7D$_{3/2}$ state differed from the
previously measured experimental value \cite{Wessel87} by ca. 15\% while the theoretical
prediction of \cite{Wijngaarden94} was lower than our
measured value by some 5\%.

To increase the accuracy and reliability of experimentally measured tensor
polarizabilities, it was useful to use the calibration with respect to a level
with well established polarizability value.  This approach substantially diminished
possible errors in the determination of external electric field values.

For the 7D$_{3/2}$ and 9D$_{3/2}$ states under study, the accuracy of existing hfs constants
limited the accuracy of the tensor polarizability measurement which could be
achieved by applying electric field induced level-crossing spectroscopy.

\section{Acknowledgements}
The authors are indebted to Bruce Shore for numerous stimulating discussions. The authors express sincere
gratitude to S. Svanberg, E. Arimondo, H. Metcalf, and V.~A. Dzuba for sharing useful information, as well
as to Janis Alnis for helping to assemble the diode lasers and to Robert Kalendarev for producing the glass
cells.
We gratefully acknowledge support from the NATO SfP 978029 Optical Field Mapping grant, from the EC 5th
Frame “Competitive and Sustainable Growth” grant G1MA-CT-2002-04063, and from the Latvian State Research Programme funding
(grant 1-23/50).
K.B., F.G. and A.J. are grateful to the European Social Fund for support.

\end{document}